\documentclass[11pt,twoside]{article}
\usepackage{asp2004}
\usepackage{psfig}
\usepackage{epsf}
\usepackage{graphics}
\usepackage{lscape}
\def\edcomment#1{\iffalse\marginpar{\raggedright\sl#1\/}\else\relax\fi}
\reversemarginpar

\begin{document}
\title{The Impact of Supernovae of Type Ia on Large Scales through the Secular Evolution of their Rate}
\author{L. Greggio}
\affil{INAF, Osservatorio Astronomico di Padova, Vicolo dell'Osservatorio 5,
I-35122 Padova, Italia}

\begin{abstract}
A straightforward formalism to evaluate the impact of Type Ia Supernovae
(SNIa) on large scale astrophysical issues is presented, together with 
analytical formulations for the SNIa rate following an instantaneous burst 
of Star Formation (SF), for the variety of SNIa progenitors.
Some applications of the parametrized formalism are illustrated.
The observations seem to favour the Double Degenerate (DD) systems as SNIa 
precursors.
\end{abstract}

\section{Introduction}

Type Ia supernovae are an important source of Iron and energy to the 
interstellar medium, and for this reason the evolution of their rate 
over cosmic times has profound implications on a variety of astrophysical 
issues, including the chemical evolution of galaxies, the Fe content
of clusters of galaxies, the evolution of the gas flows in Ellipticals (Es).
For example, the delayed release of Fe from SNIa with 
respect to the prompt release of $\alpha$ elements from Type II supernovae
implies that the shorter the formation timescale 
of a stellar system, the higher the $\alpha$/Fe abundance ratio 
recorded in its stars.
Following this argument, short formation timescales have been inferred
for E galaxies from the supersolar $\alpha$ to Fe ratios 
traced by the Mg and Fe indices in their spectra (e.g. 
Matteucci 1994). 
As another example, the interplay between the rate of mass 
return from evolved stars in Es and the rate at which this 
matter is heated by SNIa 
explosions crucially determines the dynamical evolution of the gas in these 
galaxies (Renzini 1996), i.e. whether from inflow to outflow or 
from outfow to inflow.

Our ability to address these (and other) issues is hampered by the 
uncertainty which still affects the SNIa progenitors. 
Current models include Single Degenerate (SD) 
and Double Degenerate (DD) systems, both dealing with a close binary 
in which the primary component ($m_1$) evolves into a CO white dwarf (WD). 
When the secondary ($m_2$) evolves off the MS, Roche Lobe Overflow (RLO) 
occurs: in the SD model, the WD accretes and grows in mass; in the DD model, 
the envelope
of the secondary is lost from the system, leaving a close double
WD, which eventually merge due to the emission of gravitational radiation.
In both cases, explosion may occur either when the accreting WD reaches
the Chandrasekhar limit and C deflagrates (Ch), or when a  
He layer of $\sim$ 0.15 $M_\odot$ has accumulated on top of the CO WD so 
that He detonates (S-Ch) (Hillebrandt \& Niemeyer 2000). 
The various models correspond to markedly different secular 
evolution of the SNIa rate, with great variance of 
the consequences on related issues.


To assess the impact of the SNIa rate on any astrophysical problem
it is important to (i) adequately characterize the progenitor model and
(ii) correctly couple this model to the SF history of
the specific system.
In this paper I will go through these steps, and show how 
the scenarios for the SNIa progenitors can be constrained by their 
impact on large scales. To do that I will use 
analytical relations for the SNIa rate for both SD and DDs derived in 
Greggio (2004), as a convenient parametrization of the progenitors'
models. 

\section {The Model}

An effective formulation of the SNIa rate for the modelling
of stellar systems rests upon the definition
of two key quantitites: the distribution function of the delay times 
$f_{\rm Ia}(\tau)$, where $\tau$ is the time 
between the birth of a star and its death as a SNIa, and the number fraction 
of SNIa events out of one stellar generation, $A_{\rm Ia}$. 
Denoting with $\tau_{\rm i}$ and
$\tau_{\rm x}$ the minimum and maximum delay time, and
if $A_{\rm Ia}$ is constant, the SNIa rate
at epoch $t$ in a system with a SF rate $\psi$ is:

\begin{equation}
\dot{n}_{\rm Ia}(t) = k_{\alpha} \cdot A_{\rm Ia} \cdot \int_{\tau_{\rm i}}
^{\min(t,\tau_{\rm x})}{\psi(t-\tau) \cdot  f_{\rm Ia}(\tau) \,\, {\rm d} \tau}
\label {eq_rate}
\end{equation}

\noindent
where $k_{\alpha}$ is the number of stars per unit mass in each stellar 
generation (e.g. $k_\alpha= 2.83$ for Salpeter IMF between 0.1 and 120 
$M_\odot$), considered constant in time. 





Eq.(\ref{eq_rate}) gives the dependence of the observed SNIa rate on the SF
history:  for systems with an almost constant $\psi$,
like late type galaxies, at late epochs ($t \simeq \tau_{\rm x}$):

\begin{equation}
\dot{n}_{\rm Ia}^{\rm L}(t) \simeq \langle \psi \rangle \cdot k_{\alpha} 
\cdot A_{\rm Ia} 
\label{eq_nial}
\end{equation}

\noindent
since $f_{\rm Ia}(\tau)$ is normalized to one. Thus,
the current SNIa rate in late type galaxies essentially gives information 
on the realization probability of the SNIa scenario, since it results from the
contribution of systems with all possible delay times.
By approximating $\langle \psi \rangle$ with the ratio of the galaxy mass
over its age, and assuming an $M_{\star}/L_{\rm B}$ ratio of the order 
of unity, Eq.(\ref{eq_nial}) requires $A_{\rm Ia} \sim 10^{-3}$ to match the 
observed value of 0.2 SNUs (Cappellaro, Evans, \& Turatto 1999).

In early type galaxies, where most SF has occurred in
an initial burst of relatively short duration $\Delta t$, 
the late epoch SNIa rate is:

\begin{equation}
\dot{n}_{\rm Ia}^{\rm E}(t) = M_{\star} \cdot k_{\alpha} \cdot 
A_{\rm Ia} \cdot \langle f_{\rm Ia} \rangle_{t-\Delta t,t}
\label{eq_niae}
\end{equation}

\noindent
i.e. it is proportional to the value of $f_{\rm Ia}$ at delay times 
close to the age of the galaxy, and to the total mass of stars {\em formed}
in the burst ($M_{\star}$).

If $A_{\rm Ia}$ and $k_{\alpha}$ are the same
in early and late type galaxies, the ratio between the current SNIa rates 
measured in SNU in the two galaxy types is:

\begin{equation}
\mathcal{R}_{\rm SNU} =
\frac{{\dot{n}^{\rm E}}_{Ia,SNU}}{{\dot{n}^{\rm L}}_{Ia,SNU}} \simeq 
\frac {(M_\star/L_{\rm B})^E} {(M_\star/L_{\rm B})^L} \times 
\frac {\langle f_{\rm Ia} \rangle_{t-\Delta t,t}}
{\langle f_{\rm Ia}\rangle_{\tau_{\rm i},t}}.
\label{eq_rsnu}
\end{equation}

\noindent
Using Maraston (1998) solar metallicity models, 
at an age of 12 Gyr a system formed at a 
constant rate has  $M_\star/L_{\rm B} \simeq  2$, while a single burst stellar 
population has $M_\star/L_{\rm B} \simeq 13$. 
Since $\mathcal{R}_{\rm SNU} \sim 1$ (Cappellaro et al. 1999), 
Eq. (\ref{eq_rsnu}) implies: 

\begin{equation}
\frac {\langle f_{\rm Ia} \rangle_{t-\Delta t,t}}
{\langle f_{\rm Ia}\rangle_{\tau_{\rm i},t}} \sim  0.15
\label{eq_frat}
\end{equation}

\noindent
showing that, on the average, the distribution of the delay times
must be decreasing with $\tau$, i.e.   
{\em the majority of SNIa precursors are relatively short lived}.


\vspace {0.3cm}

The theoretical SNIa rate can be obtained with numerical simulations,
which follow
the evolution of a population of binary systems, under some prescriptions 
concerning the outcome of the mass transfer phases (e.g. Han 1998).
Typically, the predicted $A_{\rm Ia}$ ranges within $10^{-4}$ and $10^{-3}$, 
while $f_{\rm Ia}$ is charaterized by an early steep rise, followed by
a decline at later epochs.
These results depend on a variety of input parameters, whose role is 
difficult to gauge, and therefore they are hardly suitable for a thorough 
exploration of the parameter space. 
On the other hand, the shape of the $f_{\rm Ia}(\tau)$ function  can be
characterized on the basis of general considerations founded on stellar
evolution. Along this line, Greggio (2004) derives analytical    
relations for the $f_{\rm Ia}$ function for 
both the SD and the DD model, briefly described below.

\begin{figure}[!ht]
\plottwo{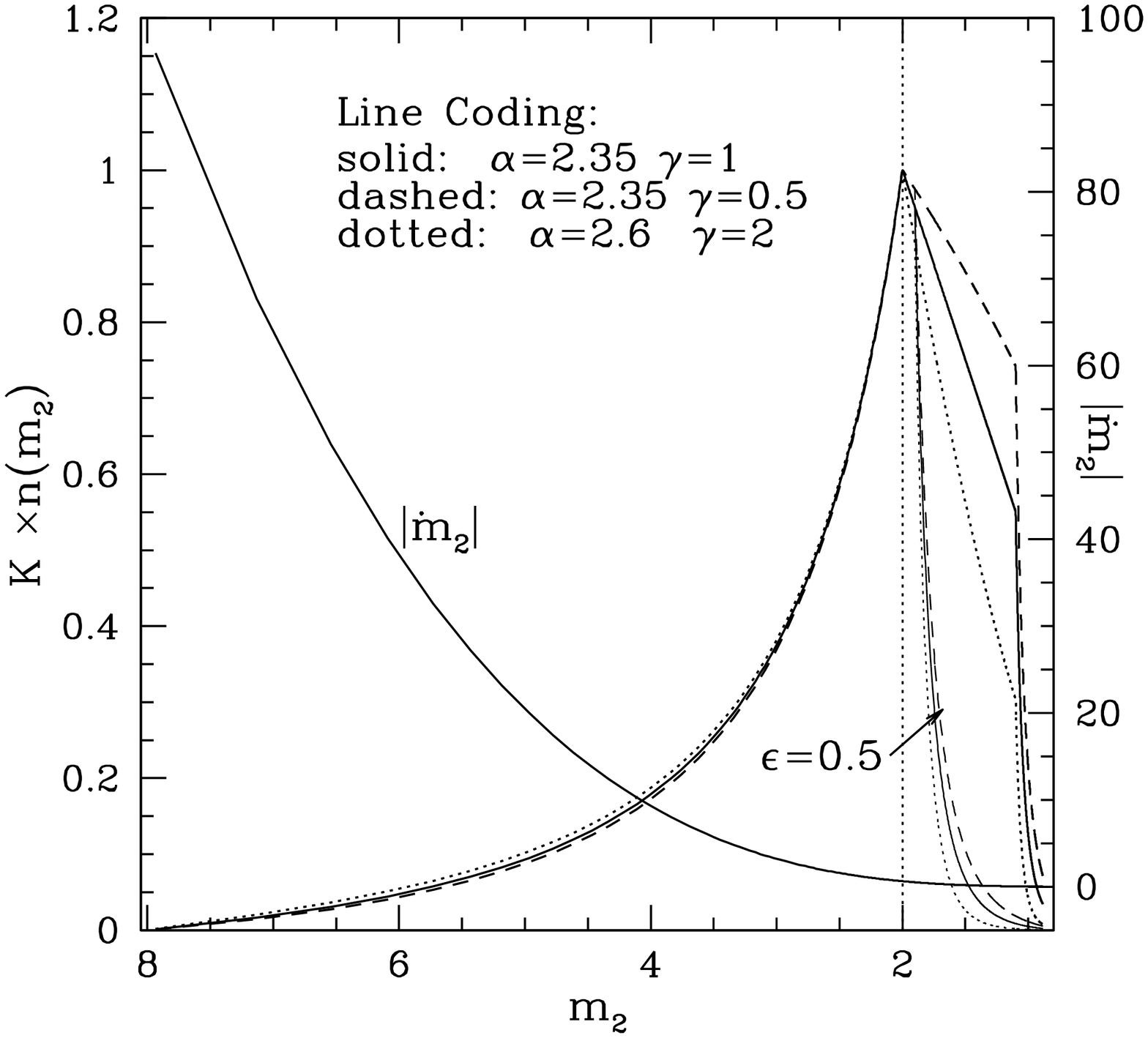}{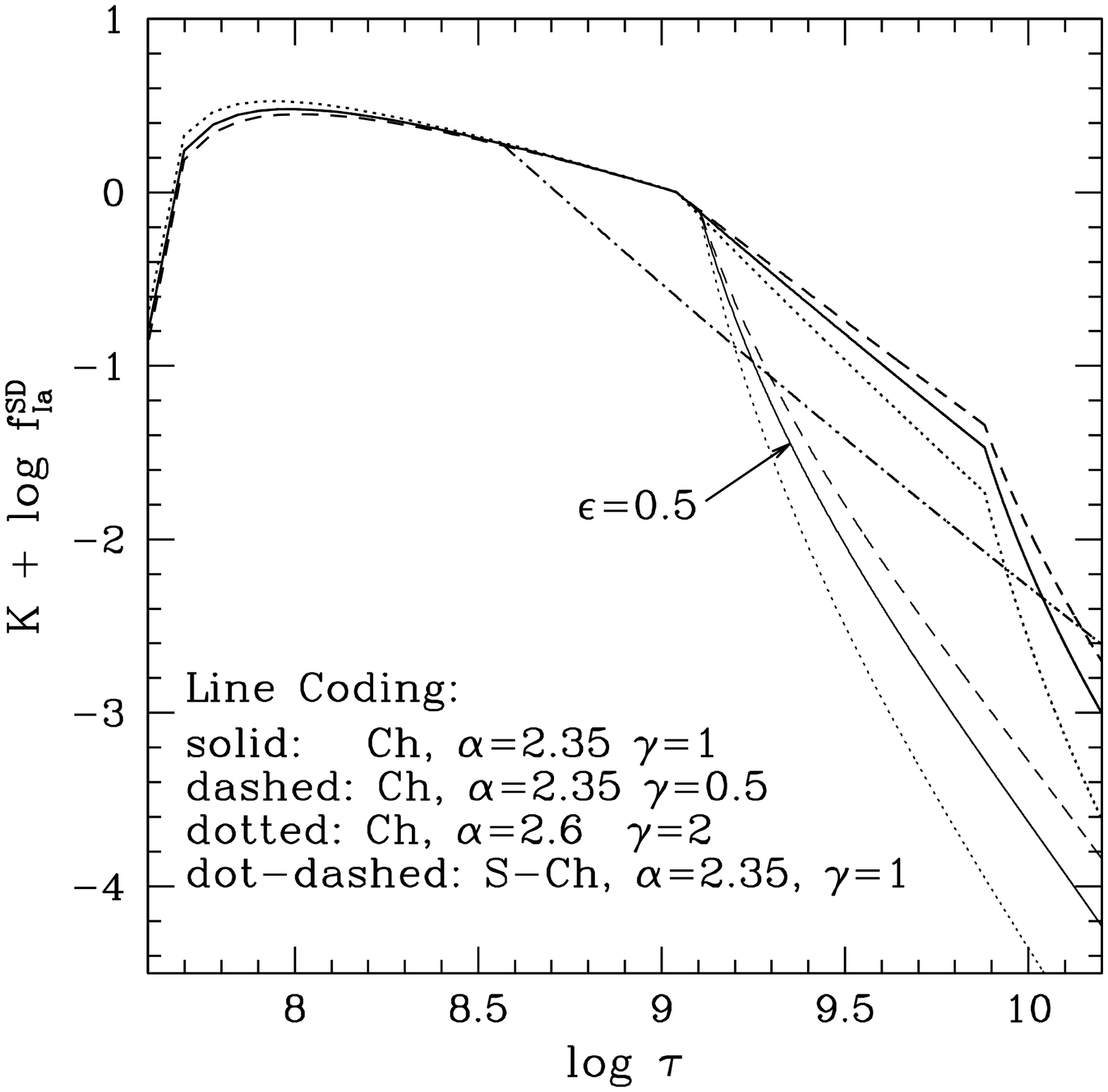}
\caption{The SD model: $n(m_{2})$ is derived from the
distributions $n(m_1) \propto m_{1}^{-\alpha}$ and $n(q) \propto q^\gamma$
where $q=m_2/m_1$, and summing over the relevant combinations. 
The right panel shows the $f_{\rm Ia}^{SD}$ function on arbitrary units, 
as from Eq. (\ref{eq_fiasd}). $\tau$ is the MS lifetime of the mass $m_2$.
$\epsilon$ is the ratio between the accreted mass and the envelope mass of the
donor. Curves are shown for $\epsilon=1$ and $\epsilon=0.5$ (labelled).
For the Sub-Ch models, it is assumed that $m_{1} \geq 3$.}

\label{fig1}
\end{figure}

\begin{figure}[!ht]
\plottwo{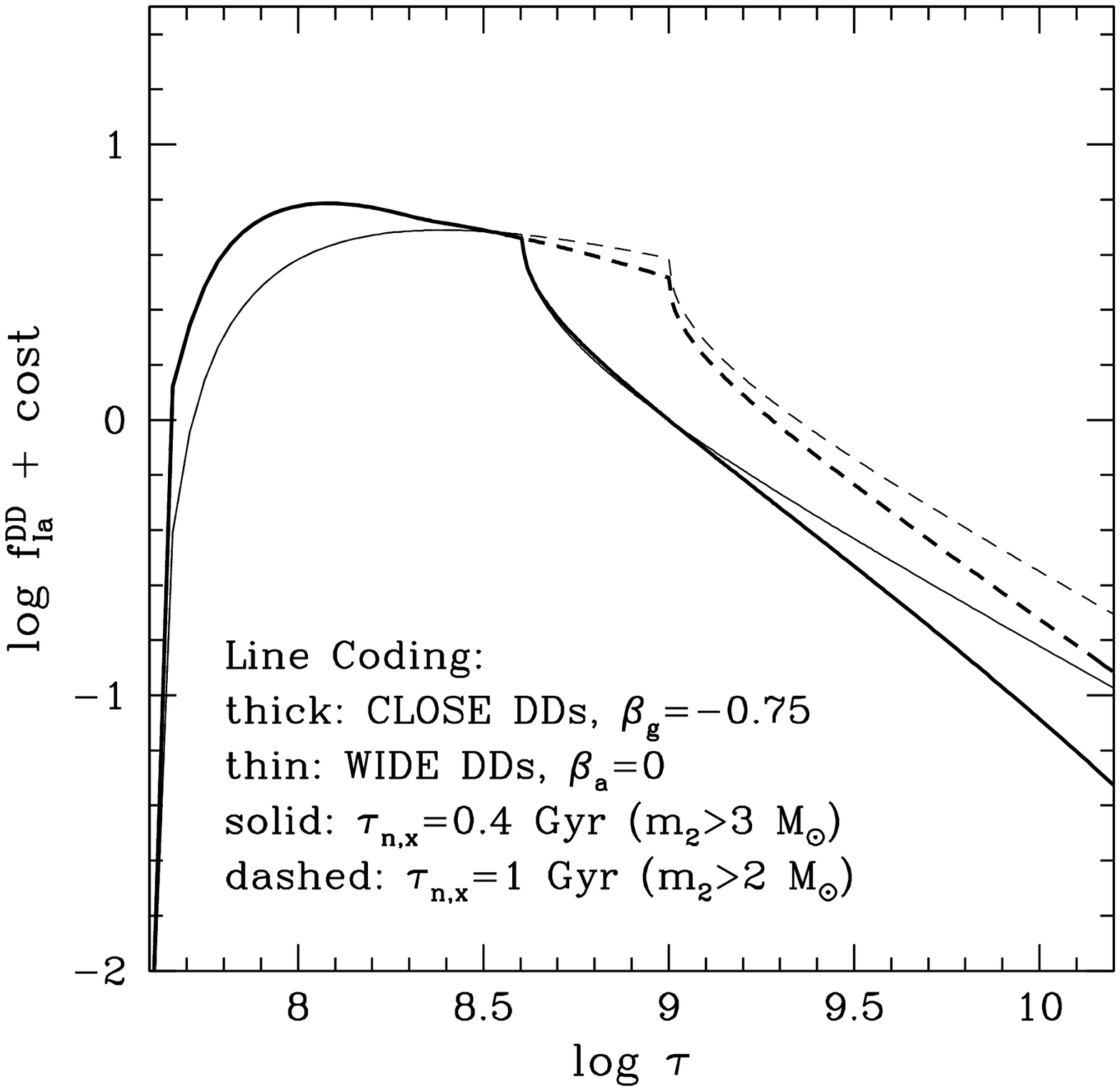}{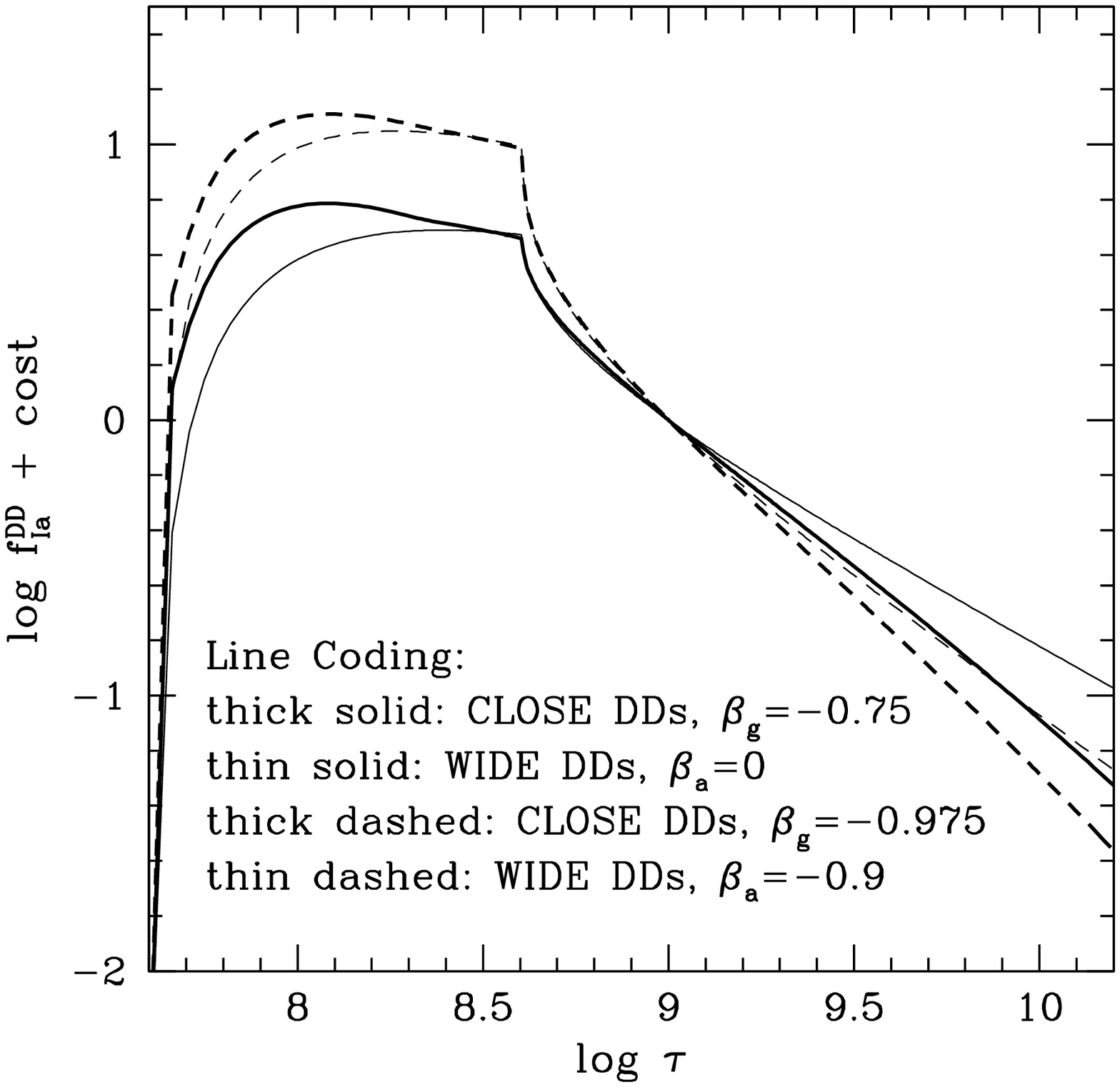}
\caption{The DD model: dependence on $\tau_{\rm n,x}$ (left), and on
the slope of the distribution of the gravitational delays (right). 
See text for more details.}
\label{fig2}
\end{figure}

In the SD model, the delay time is equal to the MS 
lifetime of the secondary, plus the duration of the accretion phase, 
which is negligible. Thus: 

\begin{equation}
f_{\rm Ia}^{SD}(\tau) \propto n(m_{2}) \cdot |\dot{m_{2}}|.
\label{eq_fiasd}
\end{equation}
 
\noindent
where $|\dot{m_{2}}|$ is the rate of change of the mass whose MS lifetime
is $\tau$, and $n(m_{2})$ is the distribution function of the secondary 
masses {\em in those systems which produce SNIa events}.
Fig. \ref{fig1} shows the two factors in Eq.(\ref{eq_fiasd}) (left panel),
and their product (right panel). The shape of the $n(m_2)$ function reflects 
the limitations imposed on the components masses in order to 
give rise to a SNIa. It is assumed that $m_1$ varies between 2 and 8 $M_\odot$,
since most CO WD come from this range. The lower boundary originates 
a relative lack of systems once $m_2$ drops below 2 $M_\odot$.
The last steep decrease, instead, reflects the need of building a WD up to
the Ch limit: as $m_2$ decreases, so does its envelope, that is
the donated mass. Then, at late epochs, only the most massive WD will 
end up as SNIa, which implies a lower limit of $m_1 > 2 M_\odot$.
Below $m_2=2 M_\odot$, $n(m_2)$ is very sensitive to 
$\alpha, \gamma, \epsilon$ (defined in the caption), 
which determine the volume in the parameter space of SNIa progenitors.
The resulting $f_{\rm Ia}$ function in shown in the right panel: notice the
dramatic drop at late epochs which stems from the requirement of building up to
the Ch mass. S-Ch exploders (dot dashed line) do not show this drop, the
only requirement being that the CO WD is more massive than 0.7 $M_\odot$. 
If additional limitations are imposed on $m_2$, 
the $f_{\rm Ia}$ function is 
non zero only in the range of $\tau$ equal to the MS lifetimes
of the secondaries with appropriate mass. 
For example, if systems with $m_2 < 2 M_\odot$ evolve into a common 
envelope, $f_{\rm Ia}^{\rm SD}=0$ for $\tau > 1$ Gyr. 
On the other hand, in order to explain the occurrence of SNIa in Es 
with the SD model, low mass secondaries must be allowed for.
  
In the DD model, the delay time is 
equal to the MS lifetime of the secondary ($\tau_{\rm n}$),
plus the gravitational delay ($\tau_{\rm gw}$). Considering only CO DDs,
$m_2$ is (in most cases) heavier than 2 $M_\odot$, which implies
$\tau_{\rm n} \leq 1$ Gyr; $\tau_{\rm gw}$ instead
can span a very wide range, being extremely sensitive to the separation $A$ 
of the DD. For example, for a (0.7+0.7) $M_\odot$ system, $\tau_{\rm gw}$ 
goes from 14 Myr to 18 Gyr when $A$ increases from 0.5 to 3 $R_\odot$. 
Since the separation of the progenitors of the CO DDs
ranges from some tens to some hundreds $R_\odot$, a great degree of
shrinkage must occur at the RLO phases in order to produce explosions
within a Hubble time.
$f_{\rm Ia}^{\rm DD}$ can be derived as a modification of 
$f_{\rm Ia}^{\rm SD}$, with early explosions from systems with short
$\tau_{\rm n}$ {\em AND} $\tau_{\rm gw}$, and late explosions from systems 
with long $\tau_{\rm gw}$. The distribution of $\tau_{\rm gw}$ will
be very sensitive to the distribution of $A$ within the range of a few 
$R_\odot$. Motivated by literature results, 
Greggio (2004) considers two possibilities: 
(i) the WIDE DDs, in which the
first RLO corresponds to a modest shrink of the system (as, e.g., in Nelemans
et al. 2001); (ii) the CLOSE DDs, similar to the classical 
recepy (as, e.g., in Han 1998) . The two channels imply 
different characterizations of the distribution of the gravitational
delays.
 Fig. \ref{fig2} shows the resulting $f_{\rm Ia}^{\rm DD}$ 
functions for some choice of the  major  
parameters: $\tau_{\rm n,x}$, the MS lifetime of the least massive secondary 
still progenitor of a SNIa; $\beta_{\rm a}$ and $\beta_{\rm g}$ which are the 
exponents of the two power laws assumed to represent the distribution of $A$ 
for the WIDE DDs, and of $\tau_{\rm gw}$ for the CLOSE DDs 
\footnote{ If SNIa originate from DD systems with approximately the same mass, 
$\beta_{\rm g} \simeq -0.75+0.25\beta_{\rm a}$}.
 
The $f_{Ia}^{DD}$ function shows an early rise, a wide maximum
and a late epoch decline. The width of the maximum is $\simeq \tau_{\rm n,x}$.
Relative to the CLOSE DD, the WIDE DD
scheme of evolution leads to flatter $f_{\rm Ia}^{\rm DD}$ distributions,
and the more systems with short $\tau_{\rm gw}$ (i.e. steeper  
$\beta_{\rm a}$ or $\beta_{\rm g}$), the steeper the early rise as well
as the late epoch decline.

\section{Constraining the SNIa Progenitors and Implications}

In summary, stellar evolution arguments strongly support a  
distribution function of the delay times characterized by:

\par\noindent
$\bullet$ an early maximum, soon after the most massive SNIa progenitor 
explodes;
\par\noindent
$\bullet$ a relatively flat portion, up to a delay time equal to the MS 
lifetime of the least massive secondary in binaries which end up as SNIa;
\par\noindent
$\bullet$ a decline phase, which for the SD Ch models becomes very steep
at late epochs, due to the requirement of building up the Chandrasekhar mass. 


\begin{figure}[!ht]
\plottwo{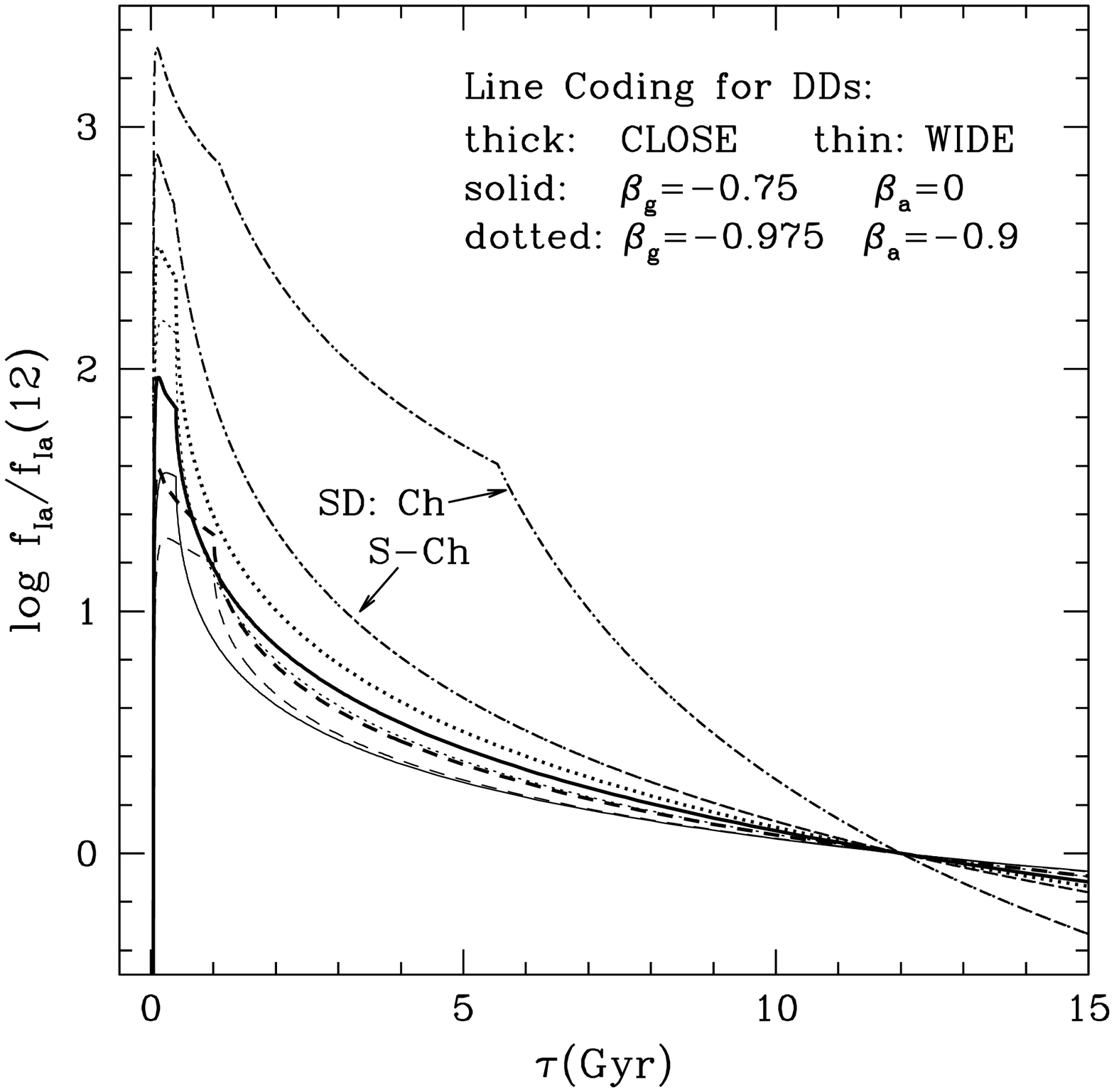}{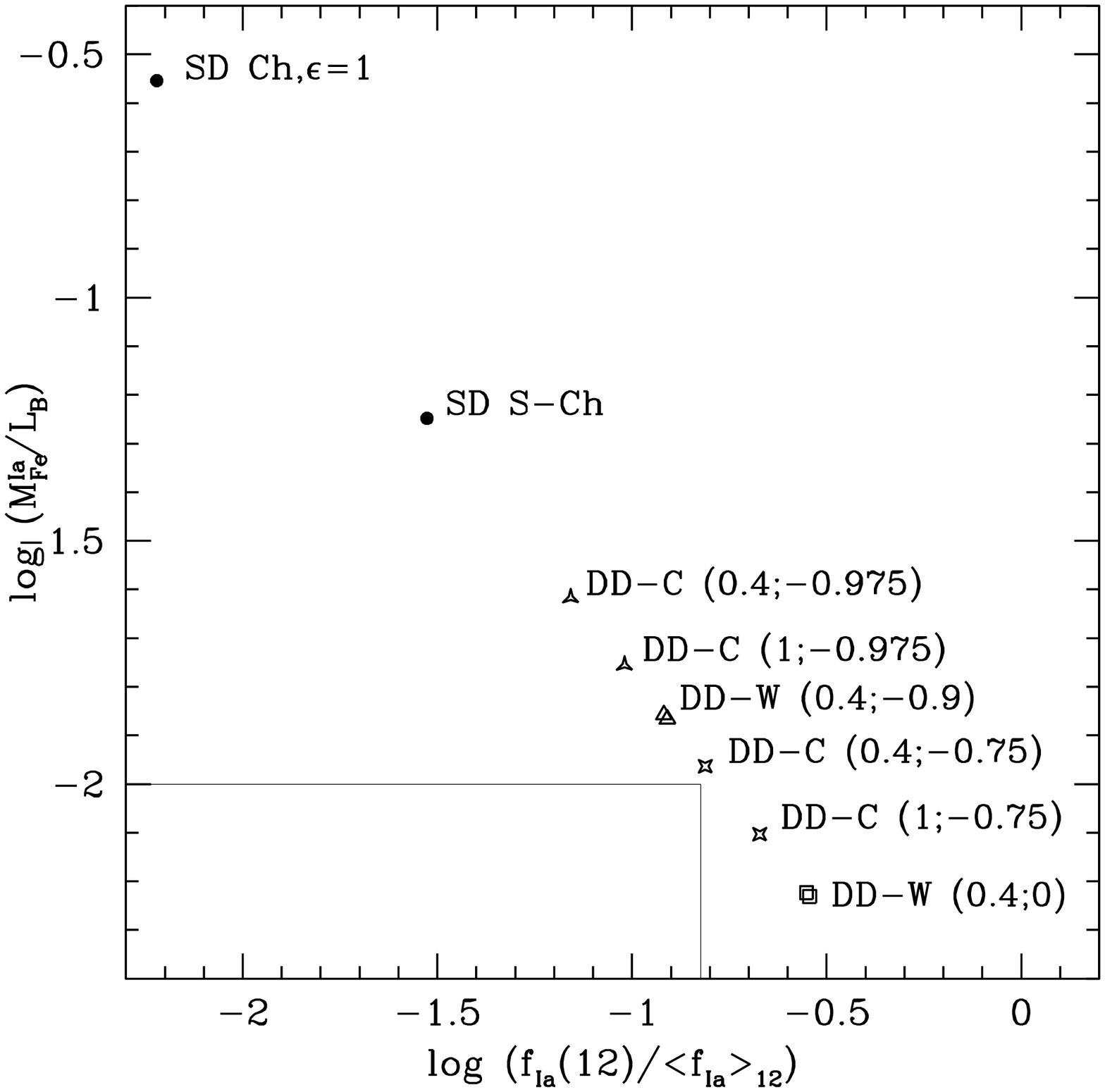}
\caption{Normalized $f_{\rm Ia}(\tau)$ functions (left) and their global 
properties (right) for the various progenitors models. 
The total Fe production per unit
luminosity of the parent galaxy is computed by adopting a current
SNIa rate in Es of 0.2 SNUs, and a release of 0.7 $M_\odot$ of Fe
per SNIa. The numbers in parenthesis are respectively $\tau_{\rm n,x}$ and
$\beta_g$ (CLOSE, DD-C) or $\beta_a$ (WIDE, DD-W). All the plotted models 
assume $\alpha=2.35,\gamma=1$ to derive the distribution of the secondary
masses in the primordial binaries.} 
\label{fig3}
\end{figure}

In spite of the similar shape, the different models have very different 
astrophysical implications: this can be appreciated by looking at the left
panel in Fig. \ref{fig3}, where the various models are plotted 
normalized at 12 Gyr, so that they all reproduce the currently observed 
SNIa rate in Ellipticals. Notice that $f_{\rm Ia}$ is proportional
to the SNIa rate following an istantaneous burst of SF, not too far from
what is expected in Es, whose stars are mostly old.
The SD Ch model predicts a huge variation of the SNIa rate, which at early
epochs would have been $\sim$ 2000 times greater than now.
Similarly, the S-Ch model corresponds to a large excursion of the SNIa rate,
but, after the maximum, the rate drops rather quickly. Finally, for
the DD models the initial peak is less strong. 

The various models are characterized by different values of 
the ratio in Eq.(\ref{eq_frat}), and they correspond to
vastly different total Fe production, as shown on the right panel of 
Fig.\ref{fig3}.
According to Renzini (2003), the Fe Mass-to-light ratio
in Clusters of Galaxies, which are dominated by Es, is $\sim$ 0.015: 
if 2/3 of this is provided by SNIa, then 
$M_{\rm Fe}^{\rm Ia}/L_{B} \sim 0.01$. 
This, admittedly rough, estimate is reported in Fig.\ref{fig3}, together
with the  empirical value for the ratio in Eq.(\ref{eq_frat}).
It appears that, once normalized to reproduce the current rate in Es,
SD models greatly
overproduce Fe in galaxy Clusters {\em AND} overpredict the current
SNIa rate in late type galaxies, where all the delay times are sampled.
The data require a less dramatic secular evolution of the SNIa rate,
similar to the behaviour of the DD models.

The different models in Fig. \ref{fig3} imply different 
timescales over which, following a burst of SF, 1/2 of the total SNIa 
explosions occur. Although the DD model is characterized by a less dramatic
evolution of the SNIa rate, since most of the explosions take place within 
the first Gyr, supersolar $\alpha$/Fe ratios still imply SF 
timescales within 1-2 Gyr.
Regarding the dynamical evolution of the gas in Es, detailed
modelling is needed, since at late epochs the SNIa rate from DDs and the rate 
of mass return decline at a similar pace ($\propto \tau^{-1.3}$).

 
\vspace{0.3cm}

In summary, taking advantage of an analytical fomulation of the distribution 
of the delay times of SNIa progenitors, I have outlined a straighforward
way to estimate their impact on the large scales. A more detailed elaboration 
of the issues considered here, as well as other applications, will be 
presented elsewhere.

\end{document}